\documentstyle[aps]{revtex} 
\begin{document}
\draft
\title{The gap maximum of anisotropic superconductors}
\author{R. Combescot}
\address{Laboratoire de Physique Statistique de 
l'Ecole Normale Sup\'erieure, associ\'e au CNRS et 
aux Universit\'es Paris 6 et Paris 7, \linebreak
24 rue Lhomond, 75005 Paris , France}
\date{Received \today}
\maketitle
\begin{abstract}
For a completely general anisotropic order parameter (including  
changes of sign), we show that weak coupling theory is incompatible  
with high values of the maximum $\Delta _{M} $  of the zero temperature  
gap as compared to the critical temperature $T_{c } $  , such as  
those found experimentally in $Bi_{2}Sr_{2}CaCu_{2}O_{8+\delta } $  
where 2 $\Delta _{M} $  / $T_{c } $  $\approx  $ 7.5 . 
This gives evidence for strong coupling effects. In particular 
this comes as a major  
support for a spin fluctuation mechanism with strong coupling,  
if one assumes that only a repulsive pairing interaction is  
at work in high $T_{c } $ superconductors. 
\end{abstract}
The last few years have seen a very important progress in the  
identification of the order parameter in some high $T_{c } $  
superconducting   
compounds. Indeed there are now quite firm experimental evidences  
that, in $YBa_{2}Cu_{3}O_{7} $ (YBCO),  it changes sign and  
that there are nodes in the gap \cite{leggett}. More recently  
a linear dependence of the penetration depth has been observed  
in $Bi_{2}Sr_{2}CaCu_{2}O_{8+\delta } $ (BSCCO) \cite{jacobs,lee},
and phase sensitive experiments have also given a positive answer  
\cite{kirtley}. There is also clear indication from Raman scattering  
experiments \cite{sacuto} of nodes in the gap in Hg compounds.  
Finally a spontaneous half magnetic flux quantum has been observed  
quite recently \cite{tsuei} in $Tl_{2}Ba_{2}CuO_{6+\delta } $  giving  
evidence for a change of sign of the order parameter. These  
features of the order parameter are a clear indication that there is 
some important repulsive contribution in the pairing interaction. The 
most obvious origin for such a contribution is Coulomb repulsion.\par
\bigskip
Yet this answer does not provide a complete physical picture  
for this repulsive component. Indeed \cite{6} this repulsion  
can appear in a direct way, or it can also be the microscopic  
origin of low energy antiferromagnetic fluctuations, with pairing  
mostly due to exchange of these fluctuations between electrons.  
A qualitative difference between these two mechanisms is the  
characteristic energy of the pairing interaction. If we are  
dealing with direct Coulomb repulsion, the typical energy entering  
the interaction is of order of Coulomb interaction itself, that  
is typically a few eV. Since the critical temperature and the  
gap are quite small compared to this energy, the pairing interaction  
can be considered as instantaneous. This implies that pairing  
can be very well described by weak coupling BCS theory. On the  
other hand if pairing is due to the exchange of spin fluctuations,  
our characteristic energy is of the order of a spin fluctuation  
frequency, which is a few tenths of eV at most. In this case  
the critical temperature and the gap are no longer small compared  
to this energy, and the pairing interaction can not be considered  
as instantaneous. This means in particular that pairing has  
to be described by a strong coupling generalization of BCS theory  
\cite{7}. Therefore we can obtain an indication on the kind  
of repulsive interaction we are dealing with by checking, as  
well as we can, if the superconductor is satisfactorily described  
by weak coupling theory or if there is a need for strong coupling effects.\par
\bigskip
As it is well known the consequences of weak coupling theory  
are much more restrictive than those of strong coupling. Therefore  
in this paper we will consider some consequence of weak coupling  
theory and see if it can be made to agree with experiments.  
Specifically we will deal with the zero temperature gap to critical  
temperature ratio. Naturally it is well known that, for isotropic  
pairing, this ratio is given by the famous BCS value 1.76 .  
However in this paper we will consider the much more general  
case of anisotropic weak coupling BCS theory for which no such  
a simple result exists. Actually, as we already mentionned,  
we can have nodes in the order parameter and in this case there  
is, strictly speaking, no gap at all for the whole excitation  
spectrum. On the other hand for a fixed value of the wavevector  
{\bf k} of the excitation, 
we have a gap $|\Delta_{\bf k}| $ for the excitation   
energy where $\Delta_{\bf k}$  is the order parameter. We are interested  
in the maximum $\Delta _{M} $  of this gap over the Fermi surface,  
and its ratio $\Delta _{M} $  / $T_{c } $ to the critical temperature.  
Our reason for investigating this ratio is that we have some  
good experimental data for it. Surprisingly this is not so much  
for YBCO, the most investigated high $T_{c } $ superconductor,  
where the data are not very clear although 2 $\Delta _{M} $  / $T_{c } $  
seems to range from 6 to 8 in most measurements. The clearest  
data are perhaps found in BSSCO where tunneling experiments  
perpendicular to the c axis \cite{renner} give a fairly sharp  
peak around 30 meV leading to 2 $\Delta _{M} $  / $T_{c } $  $\approx  $ 7.5 . 
The sharpness of the peak makes unlikely a shift to higher energy  
due to broadening. Angular dependent tunneling experiments in  
the a-b plane \cite{kane} give even a gap maximum reaching 40  
meV. One may still worry that tunneling sees only a surface  
feature. However Raman scattering clearly samples the bulk,  
and it gives \cite{staufer,dever} a maximum around 550 $cm^{-1}, $  
which seems to confirm 2 $\Delta _{M} $  / $T_{c } $  $\approx  $ 7.5 , 
although the data are broad which makes the interpretation less secure.  
Taken together these experiments are suggestive of a fairly  
high ratio. Raman data in Hg-1212 give similar results \cite{sacuto}.  
Facing these experimental data, it is of interest to investigate  
if they can be explained by taking into account the anisotropy  
of the gap within weak coupling theory. This is the purpose  
of the present paper.  \par
\bigskip 
In weak coupling theory the order parameter $\Delta_{\bf k}$  is  
obtained by solving the gap equation :
\begin{eqnarray}
{\rm \Delta }_{\bf k\rm }=\rm \int{d}_{2}\bf k\rm ' {V}_{\bf k\rm ,
\bf k\rm '}{\Delta }_{\bf k\rm '} 2\pi T 
\sum_{m=0}^{{\omega }_{c}/2\pi T}({\omega }_{m}^{2}+|{\Delta }_{\bf k'\rm }
{|}^{2}{)}^{-1/2}
\label{eq1}
\end{eqnarray} 
Here the integration $d_{2}{\bf k'}$ over the Fermi surface  
is weighted by the local density of states $[(2\pi )^{3} $  
$v_{k'}]^{-1} $  . The summation over the Matsubara  
frequencies $\omega _{m } $ = $(2m+1)\pi T $ 
is limited to a cut-off $\omega _{c } $   
large compared to the maximum $\Delta _{M} $  
of $|\Delta_{\bf k}| $  (and therefore   
to $T_{c }). $ We make no assumption on the effective interaction  
$V_{\bf k,\bf k'}$ so the situation we consider is completely  
general. We note also that a multiband model can be considered  
as a particular case of gap anisotropy so this kind of situation  
is included in our study.  \par
\bigskip 
At the critical temperature $T_{c } $  , this equation becomes  
linear. We call $\Delta_{0,\bf  k}$  the normalized eigenvector  
of $V_{\bf k,\bf k'}$ corresponding to the largest eigenvalue  
$\lambda _{0}. $ It gives the shape of the gap at $T_{c } $  . Making use  
of  $2\pi T_{c } $ $\Sigma  $ $1/|\omega _{m }| $  
= $ \ln $ ( 1.13 $\omega _{c } $ / $T_{c } $  
) = $1/\lambda _{0}, $ (valid in the weak coupling limit of 
large $\omega _{c } $  / $T_{c } $ ), we have :
\begin{eqnarray}
{\rm \Delta }_{0,\bf k\rm }=\rm \ln(1.13{{\omega }_{c} \over
{T}_{c}})\int{d}_{2}\bf k\rm '{V}_{\bf k\rm ,\bf k\rm '}{\Delta
}_{0,\bf k\rm '}
\label{eq2}
\end{eqnarray} 
Below $T_{c } $  , $\Delta_{\bf k}$  is obtained from Eq.(1) . Now an  
essential feature of this equation is that the sum over Matsubara  
frequencies is dominated by the terms $\omega _{m } $ $>>\Delta _{M} $  . 
This is seen by rewriting it as :
\begin{eqnarray}
{\rm \Delta }_{\bf k\rm }=\rm \ln(1.13{{\omega }_{c} \over
{T}_{c}})\int{d}_{2}\bf k\rm '{V}_{\bf k\rm ,\bf k\rm '}
{\Delta }_{\bf
k\rm '} +\int{d}_{2}\bf k\rm '{V}_{\bf k\rm ,\bf k\rm '}{\Delta
}_{\bf k\rm '}
\rm [2\pi T\sum_{m=0}^{{\omega }_{c}/2\pi T}
({\omega }_{m}^{2}+|{\Delta }_{\bf k\rm '}
{|}^{2}{)}^{-1/2}-\sum_{m=0}^{{\omega }_{c}/2\pi {T}_{c}}
|m+1/2{|}^{-1}]
\label{eq3}
\end{eqnarray} 
In the second term of the right-hand side we can let $\omega _{c } $  
go to infinity because the result is convergent. Since in weak  
coupling $\ln ( 1.13 \omega _{c }   / T_{c } )$ is large, we see that  
the first term dominates over the second one. Therefore to lowest  
order the shape of the gap below $T_{c } $ is still given 
by $\Delta _{0,\bf k}$. However the size of the gap is fixed 
by the second term which is non linear. 
We can obtain a still exact equation for  
this size by multiplying Eq.(3) by $\Delta ^{*}_{0,\bf  k}$  and  
integrating over {\bf k} (which takes into account the local  
density of states), leading to :
\begin{eqnarray}
\int_{}^{}{\rm d}_{2}\bf k\rm {\Delta }_{0}^{{}^*}(\bf k\rm )
{\Delta }^{}(\bf
k\rm )
\rm [\pi T\sum_{m=0}^{{\omega }_{c}/2\pi T}
({\omega }_{m}^{2}+|{\Delta }_{\bf k\rm '}
{|}^{2}{)}^{-1/2}
-\rm \sum_{m=0}^{{\omega }_{c}/2\pi {T}_{c}}
|2m+1{|}^{-1}]=0
\label{eq4}
\end{eqnarray} 
where we have made use of Eq.(2) to eliminate the interaction  
(using the fact that it is hermitian). Since this equation does  
not contain large terms anymore, we can replace $\Delta _{0,\bf k}$
by $\Delta_{\bf k}$ to lowest order. This equation can also be  
rewritten as :
\begin{eqnarray}
\rm \ln({{T}_{c} \over T})\int{d}_{2}\bf k\rm {\left|{{\Delta
}_{\bf k\rm }}\right|}^{2}=\int{d}_{2}\bf k\rm 
{\left|{{\Delta }_{\bf
k\rm }}\right|}^{2}
\rm 2\pi T\sum_{m=0}^{\infty }
\left[{{\left|{{\omega }_{m}}\right|}^{-1}-({\omega }_{m}^{2}+
|{\Delta}_{\bf k\rm }{|}^{2}{)}^{-1/2}}\right]
\label{eq5}
\end{eqnarray} 
In particular we obtain at T = 0 :
\begin{eqnarray}
\rm \ln({{\Delta }_{M} \over 1.76{T}_{c}})=-{\int{d}_{2}\bf k\rm
{\delta }^{2}(\bf k\rm )\ln(\delta (\bf k\rm )) \over \int{d}_{2}\bf
k\rm {\delta }^{2}(\bf k\rm )}
\label{eq6}
\end{eqnarray}
where we have introduced $\delta (${\bf  k}) = $|\Delta_{\bf  k}| $ /  
$\Delta _{M} $  which is the absolute value of the gap normalized to  
its maximum value. This equation  
has already been essentially obtained by Pokrovskii \cite{pokro}.
Provided that we know the shape $\delta (${\bf  k})  
of the gap, it gives us the maximum of the zero temperature  
gap $\Delta _{M} $  compared to the standard BCS value 1.76 $T_{c } $  
 . We see that the result is not sensitive to the detailed structure  
of the gap since the logarithm is a smooth function. \par
\bigskip
The above result has been obtained within weak coupling theory  
where the parameter $\omega _{c } $  / $T_{c } $  is large. If we want  
to improve on this result, we have to consider that $\omega _{c } $  
 / $T_{c } $ is not large anymore which implies to go to strong  
coupling theory anyway. There is no way to improve consistently  
on this result within weak coupling theory. Nevertheless we  
might worry that the above result is a poor approximation because  
the dominant term is only logarithmically large. Fortunately  
the general situation is much better. This can be seen by rewriting  
the exact Eq.(3) at T = 0 as :
\begin{eqnarray}
{\rm \Delta }_{\bf k}=\rm \ln(1.13{{\omega }_{c} \over
{T}_{c}})\int{d}_{2}\bf k\rm '{V}_{\bf k\rm ,\bf k\rm '}
{\Delta }_{\bf k\rm '}
+\rm \int{d}_{2}\bf k\rm '{V}_{\bf k\rm ,\bf k\rm '}
{\Delta }_{\bf k\rm '}\ln({1.76{T}_{c} \over \left|
{{\Delta }_{\bf k\rm '}}\right|})
\label{eq7}
\end{eqnarray}
and projecting it on the complete set of normalized eigenvectors  
$\Delta _{m,\bf k}$  of $V_{\bf k,\bf k'}$. If the corresponding  
eigenvalues are $\lambda _{m } $ , this gives :
\begin{eqnarray}
\rm ({1 \over {\lambda }_{m}}-{1 \over {\lambda }_{0}})\int{d}_{2}\bf
k\rm {\Delta }_{m,\bf k\rm}^{{}^*}{\Delta }_{\bf k\rm }=
\int{d}_{2}\bf
k\rm {\Delta }_{m,\bf k\rm}^{{}^*}{\Delta }_{
\bf k\rm }\ln({1.76{T}_{c}
\over \left|{{\Delta }_{\bf k\rm }}\right|})
\label{eq8}
\end{eqnarray} 
where we can think of evaluating the right-hand side to  
lowest order by replacing $\Delta_{\bf k}$ by $\Delta _{0,\bf k}$. From  
this equation, the components a $_{m } $ =  $\int  $ $\Delta _{m,\bf k}^{*}$    
$\Delta_{\bf k}$ of the gap on the eigenvectors with m $\not=  $ 0 are  
small because the eigenvalues $\lambda _{m } $  are small ( this is the  
weak coupling limit ). In addition the subdominant order parameters  
$\Delta _{m } $  will correspond  in the general case to 
critical temperatures much smaller than $T_{c }, $ which implies 
$\lambda _{m } $  $<< $ $\lambda _{0} $   
for m $\not=  $ 0 ( a specific example of this can be found in the recent  
work of Palumbo et al. \cite{palumbo} where, within a given  
channel,  $\lambda _{1} $  $\approx  $ 0.1 $\lambda _{0} $ 
is found for all the channels).   
This implies that $a_{m } $ is reduced by a 
factor $\lambda _{m } $  / $\lambda _{0} $   
with respect to a naive evaluation ( in the case of a separable  
potential one has exactly $\lambda _{m } $  = 0 and all the $a_{m } $ 's  
are exactly zero ). The opposite case 
of $\lambda _{m } $  $\approx  $ $\lambda _{0} $   
corresponds to an accidental situation and is most likely to  
be found for $\Delta _{m,\bf  k}$ and $\Delta _{0,\bf  k}$ belonging  
to different irreducible representations. This should give a  
smeared second transition below $T_{c } $ which has not been seen  
up to now in high $T_{c } $  superconductors ( except for very  
recent experiments \cite{srikanth} on the penetration depth  
in YBCO which seems to indicate the need of a multiband description;  
we will come back to this below ).  Next we see that in the  
right-hand side of Eq.(8) the logarithm will have a rather small  
absolute value in most of the range of integration, and it will  
change sign. Moreover in the cases we are interested in, $\Delta_{\bf k}$   
also changes sign, and so does $\Delta _{m,\bf  k}$ in the general  
case. Therefore we have plenty of reasons for destructive interference  
which will make the right-hand side small in general. The conclusion  
is that taking $\Delta_{\bf k}$ proportional to $\Delta _{0,\bf  k}$
is a quite good approximation. Moreover replacing in Eq.(4)  
$\Delta _{0,\bf  k}$ by $\Delta_{\bf k}$ 
should not change much the result   
when their shape is similar. Hence this should give a very good  
evaluation for the maximum $\Delta _{M} $  of the gap, which is not  
sensitive anyway to the detailed structure of $\Delta_{\bf k}$ as  
we have seen.  \par
\bigskip 
In order to calculate $\Delta _{M} $  from Eq.(6) we only need to know  
the weight function $N(\delta ) $ for the reduced gap values $\delta . $ We  
introduce the integrated weight x by  dx = $N(\delta ) $ $d\delta . $ 
In Eq.(6) we can assume by a change of variables 
that 0 $\leq  $ x $\leq  $ 1 . This gives :
\begin{eqnarray}
\rm \ln({{\Delta }_{M} \over 1.76{T}_{c}})=-{\int_{0}^{1}{dx}^{}{\delta
}^{2}(x)\ln(\delta (x)) \over \int_{0}^{1}{dx}^{}{\delta }^{2}(x)}
\label{eq9}
\end{eqnarray} 
where $\delta (x) $ is a growing function of x
with 0 $\leq  $ $\delta  $ $\leq  $ 1 . In the case of two dimensional
superconductors, which are   a very good approximation for all the known high
$T_{c } $ superconductors,   x is merely the curvilinear abscissa along the
Fermi line weighted   by the local density of states. We can check Eq.(9) in a
variety   of cases. For a constant gap, $\delta  $ = 1 and we have naturally
the   BCS result $\Delta _{M} $  / $T_{c } $  = 1.76 (with 1.76 $\equiv  $
$\pi  $   $/ $ $e^{C} $  
 where C is the Euler constant). For the A phase of superfluid  
$^{3}He, $ $\Delta _{M} $  / $T_{c } $  
= 1.76  $e^{5/6} $ /2 $\approx  $ 2.029 (  
here $\delta (x) $ = $[x(2-x)]^{1/2} $ ). For d-wave, $\delta (x) $ = sin (  
$\pi x/2 $ ) after change of variable and $\Delta _{M} $  / $T_{c } $  = 1.76   
 $2e^{-1/2} $  $\approx  $ 2.139 . For the simple model introduced by  
Xu et al. \cite{xu}, with a variable slope at the node,  $\delta (x) $  
= x / $\alpha  $  for  0 $\leq  $ x $\leq  $ $\alpha  $  
and  $\delta (x) $ = 1 for   $\alpha  $ $\leq  $ x $\leq  $  1  
. We find $\Delta _{M} $  / $T_{c } $  = 1.76  
exp ( $\alpha  $  / (9 - $6\alpha ) $ )  
which agrees with their result  $\Delta _{M} $  / $T_{c } $  = 1.994   
for their $\mu  $ $\equiv  $ 4 $/(\pi \alpha ) $ = 2 ; 
for  $\mu  $ = 2.7 we find $\Delta _{M} $  
 / $T_{c } $  = 1.904  in agreement with them ; finally for the  
upper limit  $\alpha  $ = 1  this gives the highest possible value for  
this model  $\Delta _{M} $  / $T_{c } $  = 2.462 . We note that this ratio  
is not so large, even for this model which does not look very  
physical in this limit ( at least for a single band ). It is  
then interesting to generalize this model into $\delta (x) $ = $x^{n } $  
 which gives a very wide gap opening for large n. This leads  
to $\Delta _{M} $  / $T_{c } $  = 1.76  exp ( n  / ( 2n + 1 ) ) , which  
gives 2.631 for n = 2 and saturates at  $\Delta _{M} $  / $T_{c } $  =  
1.76  exp ( 1  / 2 ) $\approx  $ 2.908 for n $\rightarrow  $ $\infty . $ 
These few examples show that it is quite hard to increase 
$\Delta _{M} $  / $T_{c } $ even by going to pretty unphysical models.  \par
\bigskip 
In order to explore more fully this question it is convenient  
to use  y = $\delta ^{2}(x) $ as a new variable and  X(y) = 1 - x   
as a new function. The graph X(y) is trivially related to the  
graph  $\delta ^{2}(x) $  and it decreases from ( y = 0 , X = 1 ) to  
( y = 0 , X = 1 ) (even when $\delta (x) $ is discontinuous, which occurs  
for example when the gap is constant ). We can then rewrite  
Eq.(9) into :
\begin{eqnarray}
\rm 2\ln({{\Delta }_{M} \over 1.76{T}_{c}})=
-1-{\int_{0}^{1}{dy}^{}X(y)\ln(y) \over \int_{0}^{1}{dy}^{}X(y)}
\label{eq10}
\end{eqnarray}
This expression Eq.(10) makes it obvious that, for fixed  
area    $\int  $ dy X(y)  =  $\int  $ dx $\delta ^{2}(x) $ , 
the maximum  $\Delta _{M} $   
/ $T_{c } $    is obtained by squeezing as much as possible the  
weight of X(y) at low  y   in order to take advantage of the  
divergence of $\ln(y)$  for y $\rightarrow  $  0. At the same 
time one sees that  
this is not very efficient in order to obtain high $\Delta _{M} $   
/ $T_{c } $    since the divergence of $\ln(y)$ is weak.  This squeezing  
is optimally reached by taking a constant gap almost everywhere  
: $\delta (x) $ = $\delta _{m } $ for 0 $< $ x $< $ 1, 
with $\delta (0) $ = 0  and $\delta (1) $  
= $1, $ equivalent to X(y) = 1 for 0 $\leq  $ y $< $ $\delta _{m }^{2} $   
and  X(y) = 0 for $\delta _{m }^{2} $   $< $ x $\leq  $ 1. 
This conclusion can also   
be found from Eq.(6) by a convexity argument , as done by Anderson  
and Morel \cite{morel}. The corresponding maximum value of the  
gap is $\Delta _{M} $  / $T_{c } $  = 1.76 / $\delta _{m } $ . 
This shows that,  by letting  $\delta _{m } $ $\rightarrow  $ 0, 
we can obtain in principle $\Delta _{M} $   
 / $T_{c } $  as high as we like. However this optimal model is  
quite unphysical since the gap maximum has zero weight, and  
is therefore irrelevant ( it will not be seen in any experiment  
). The real physical gap maximum in this model is 
$\Delta _{M} $  $\delta _{m } $   
, not $\Delta _{M} $  , and we find the BCS value for the gap to $T_{c } $  
ratio, which is expected since the gap is constant.   \par
\bigskip 
We can consider a slightly more reasonable model by giving a  
weight 1 - $x_{0} $ to the gap maximum. In order to obtain the  
optimal  $\Delta _{M} $  / $T_{c } $    we take the rest of the gap at  
a constant value $\delta _{m } $ . Explicitely this leads us to the  
simple model  $\delta (x) $ = $\delta _{m } $ for 0 $< $ x $< $ $x_{0}, $ 
and $\delta (x) $   
= $1 $  for $x_{0} $ $< $ x $\leq  $ 1. We could try to go continuously from  
$\delta _{m } $  to 1 , in order to obtain 
a better  $\Delta _{M} $  / $T_{c } $   
. However it is quite clear that, if we want that the maximum  
gets a significant weight, the improvement will be very small.  
The above model gives  $\ln r $ = - $x_{0} $  $\delta _{m }^{2} $ 
$\ln (\delta _{m } $   
 ) / $(x_{0} $  $\delta _{m }^{2} $  + 1 - $x_{0} $  ) 
with r = $\Delta _{M} $   
 / 1.76 $T_{c } $  . We can again in principle obtain  r  as high  
as we like by letting $\delta _{m } $  and 1 - $x_{0} $  go to zero.   
More generally, independently of a specific model, it is obvious  
from Eq.(10) that, in order to obtain a large $\Delta _{M} $  / $T_{c }, $  
we need qualitatively a gap maximum with a small weight and  
small gap with a large weight. However the contours of constant  
 r  of our optimal model are plotted in Fig.1 in the ( $\delta _{m } $  
, $x_{0} $  ) plane and they make quite clear quantitatively the  
difficulty which is met when one tries to obtain at the same  
time a large $\Delta _{M} $  / $T_{c } $  and a sizeable weight 1 - $x_{0} $   
for the maximum. For fixed  r , the maximum possible weight  
(1 - $x_{0})_{max} $   is given by  (1 - $x_{0})_{max} $ =  
1 / ( 1 + $2er^{2}\ln(r)) $ with a corresponding value 1 / $(r\surd e) $  
for $\delta _{m } $ . While 2 $\Delta _{M} $  / $T_{c } $ 
= 5 gives a maximum   
weight 0.2 with $\delta _{m } $ = 0.42 , the average experimental value  
2 $\Delta _{M} $  / $T_{c } $  = 7 leads to a maximum weight of 0.06 .  
This is not compatible with the experimental data, such as tunneling  
or Raman scattering which give a gap maximum with a fairly sizeable  
weight. Also the rest of the gap would be at 0.3 the maximum  
gap value which is rather low. We note that, although the above  
model is already not compatible with experiments, it does not  
even have nodes in the gap. The following model for the gap  
distribution $\delta (x) $ = (x / $x_{0}) $ $\delta _{m } $ 
for 0 $< $ x $< $ $x_{0}, $   
and $\delta (x) $ = $1 $  for $x_{0} $ $< $ x $\leq  $ 1 , 
is similar to the preceding   
one, but it is somewhat more realistic since it allows for nodes  
in the gap. It leads to  $\ln r $ = $x_{0} $  $\delta _{m }^{2} $ ( 1/3  
- $\ln (\delta _{m } $  )) / $(x_{0} $  $\delta _{m }^{2} $  
+ 3(1 - $x_{0})). $   
As expected it gives somewhat worst results for 
$\Delta _{M} $  / $T_{c }. $  \par 
\bigskip 
  Let us summarize the situation. We expect a physically reasonable  
one-band model to produce a fairly regular gap function, similar  
for example to the standard d-wave order parameter. All the  
various specific examples of this kind that we have considered  
above gave 2 $\Delta _{M} $  / $T_{c } $  scattered between 4 and 4.5  
. Values near 5 correspond already to rather unphysical situations.  
Hence weak coupling is far off the experimental result. Since  
$\Delta _{M} $  / $T_{c } $    depends on broad features of 
the gap distribution and not on details, as we have discussed, 
this result is generic not accidental. The failure to find higher 
values than, say, 5 within weak coupling theory is not due to a lack 
of inspiration in finding the proper order parameter. It is a 
systematic deep limitation of weak coupling theory itself.   \par
\bigskip 
From Eq.(10) the only way to increase $\Delta _{M} $  / $T_{c } $   within  
weak coupling is to lower the average value of the gap, while  
keeping at the same time a sizeable weight near the gap maximum  
to obtain agreement with tunneling and Raman data. This goes  
in the direction of a somewhat discontinuous order parameter  
which does not look like a simple one-band model, although one  
might argue that the spin fluctuation model with strongly peaked  
interactions at wavevector $(\pm \pi ,\pm \pi ) $ could produce such a result.  
We believe that a rather natural realization of such a strange  
gap structure is merely a two-band model (which is included  
in our study ), with one band corresponding to the maximum gap  
value and the other one to the small value. Nevertheless we  
have seen that even our optimized model can not reproduce at  
the same time the fairly large $\Delta _{M} $  / $T_{c } $    observed  
experimentally together with a reasonable weight for this gap  
maximum. We come to the conclusion that simple weak coupling  
theory is not compatible with experiment.  \par
\bigskip 
What are the ways out ? The most obvious one is to question  
experiments. As we have seen, this is not an easy way since  
independent experiments are in reasonable agreement. However  
one may wonder if tunneling or Raman experiments do not miss  
a part of the Fermi surface. In this case the weight of the  
gap maximum might be less than it seems, releasing a part of  
the theoretical constraint. The most natural situation where  
this would occur is a two-band model, where the band with the  
gap maximum would be seen but not so much the other one. On  
the theoretical side one may object that the weak coupling equation  
Eq.(1) that we have used does not include the possibility of  
a density of states varying strongly perpendicularly to the  
Fermi surface, as could be produced by nearby Van Hove quasi-singularities.  
However we know \cite{rc} that in the isotropic case we have  
quite generally  2 $\Delta _{M} $  / $T_{c } $    $\leq  $ 4  even for such a   
varying density of states, so that the prospects in this direction  
are not good. Therefore, if we stop short of rejecting BCS theory altogether,
the most likely explanation for the  
high experimental value of $\Delta _{M} $  / $T_{c } $  is that weak coupling   
theory does not apply because strong coupling effects are important.  
Indeed their existence is supported independently by various  
experiments and they are known to increase in a quite sizeable  
way this ratio. However strong coupling effects with isotropic  
pairing would have a hard time explaining  2 $\Delta _{M} $  / $T_{c } $  
 = 7 , since this would require \cite{strcpl} a quite high coupling  
constant ( at least 5 ). A possibility is to have strong coupling  
effects in a multiband model. On the other hand self-consistent  
calculations for simple d-wave pairing within the 2D Hubbard  
model have given results \cite{monthoux} as high as 2 $\Delta _{M} $  
 / $T_{c } $  $\approx  $ 10 . Hence it seems that the large experimental  
$\Delta _{M} $  / $T_{c } $  can be accounted for by strong coupling together   
with anisotropy, whereas it is incompatible with weak coupling  
theory. This comes as a strong support in favor of a spin fluctuation  
mechanism in the debate about the nature of the repulsive pairing  
interaction that we considered in the introduction.  \par
\bigskip 
We are extremely grateful to A. J. Leggett, D. Pines, D. Rainer  
and J. A. Sauls for very stimulating discussions.  \par 
\begin{figure}
\caption{ Fig. 1 Contours of constant  
r = $\Delta _{M} $  / 1.76 $T_{c } $     
in the  $\delta _{m } $ - $x_{0} $    plane for the 
model $\delta (x) $ = $\delta _{m } $   
for $ 0 <  x <  x_{0}$, and $\delta (x) $ = 1   
for $x_{0} <  x \leq  1 $. The  
values of r are 1.1 , 1.2 , 1.3 , 1.5 , 1.7 , 2. and 2.5 as  
indicated near the curves.}
\label{Fig1}
\end{figure}
%
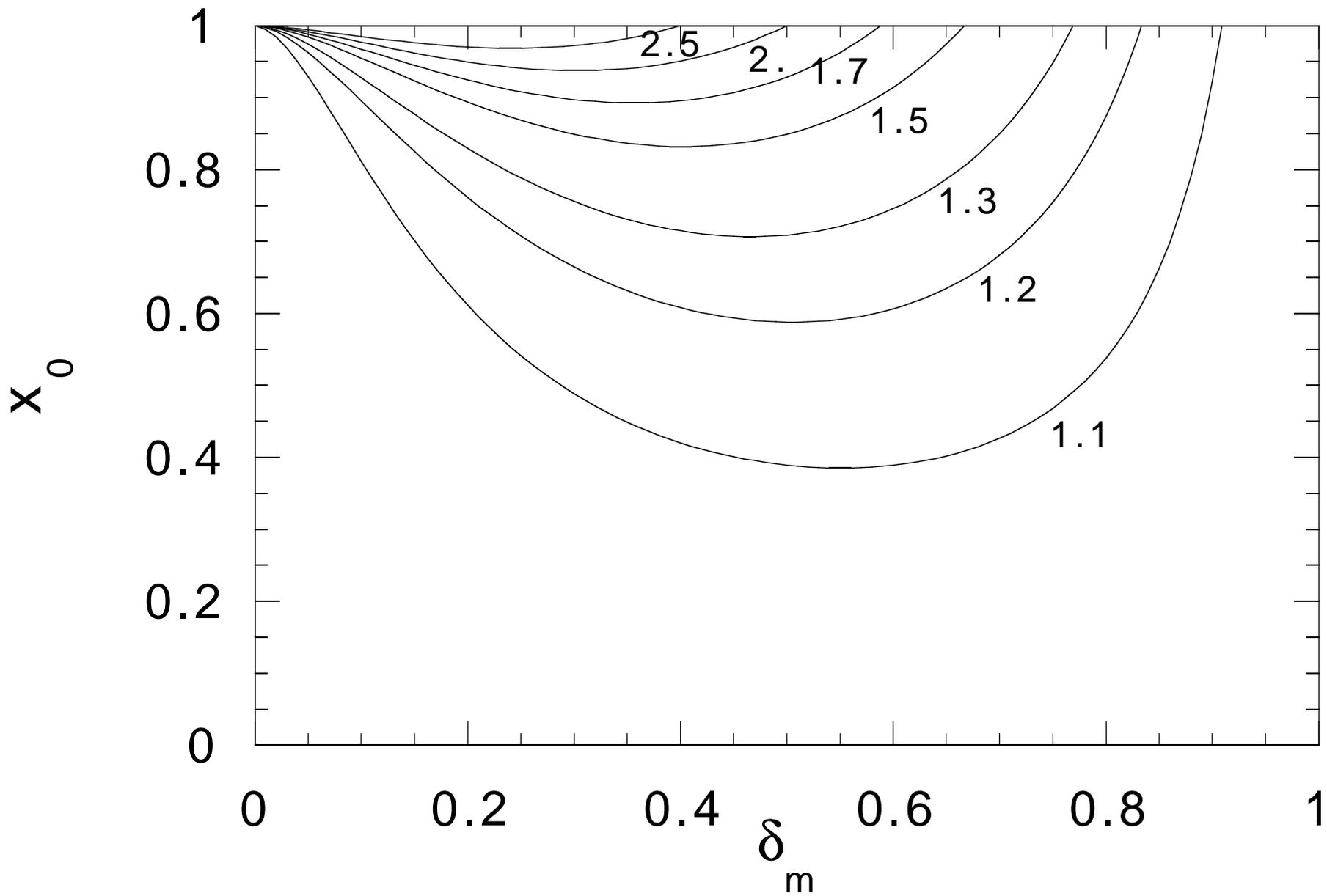
\end{document}